\documentclass[aps,prl,twocolumn,superscriptaddress,floatfix,showpacs]{revtex4-1}

\usepackage{graphicx,bm,amssymb,amsmath,dcolumn,hyperref}
\usepackage{multirow}
\usepackage{enumerate}
\usepackage{enumitem}
\usepackage{color}
\usepackage{subfigure}  
\usepackage{epstopdf}
\usepackage{bbm}
\usepackage{graphicx}
\usepackage{hyperref}
\usepackage{threeparttable}
\usepackage{amsthm}
\usepackage{mathtools}
\usepackage{color}
\usepackage{tikz}
\usepackage{float}
\usepackage{empheq}

\begin{document}
	\newcommand{\upcite}[1]{\textsuperscript{\textsuperscript{\cite{#1}}}}
	\newcommand{\be}{\begin{equation}}
		\newcommand{\ee}{\end{equation}}
	\newcommand{\half}{\frac{1}{2}}
	\newcommand{\ith}{^{(i)}}
	\newcommand{\im}{^{(i-1)}}
	\newcommand{\gae}
	{\,\hbox{\lower0.5ex\hbox{$\sim$}\llap{\raise0.5ex\hbox{$>$}}}\,}
	\newcommand{\lae}
	{\,\hbox{\lower0.5ex\hbox{$\sim$}\llap{\raise0.5ex\hbox{$<$}}}\,}
	
	\definecolor{blue}{rgb}{0,0,1}
	\definecolor{red}{rgb}{1,0,0}
	\definecolor{green}{rgb}{0,1,0}
	\newcommand{\blue}[1]{\textcolor{blue}{#1}}
	\newcommand{\red}[1]{\textcolor{red}{#1}}
	\newcommand{\green}[1]{\textcolor{green}{#1}}
	\newcommand{\orange}[1]{\textcolor{orange}{#1}}
	\newcommand{\yd}[1]{\textcolor{blue}{#1}}
	
	\newcommand{\scrA}{{\mathcal A}}
	\newcommand{\scrE}{{\mathcal E}} 
	\newcommand{\scrF}{{\mathcal F}} 
	\newcommand{\scrL}{{\mathcal L}}
	\newcommand{\scrM}{{\mathcal M}} 
	\newcommand{\scrN}{{\mathcal N}}
	\newcommand{\scrS}{{\mathcal S}}
	\newcommand{\scrs}{{\mathcal s}}
	\newcommand{\scrP}{{\mathcal P}}
	\newcommand{\scrO}{{\mathcal O}}
	\newcommand{\scrR}{{\mathcal R}}
	\newcommand{\scrC}{{\mathcal C}}
	\newcommand{\scrV}{{\mathcal V}}
	\newcommand{\scrD}{{\mathcal D}}
	\newcommand{\scrG}{{\mathcal G}}
	\newcommand{\scrW}{{\mathcal W}} 
	\newcommand{\scrU}{{\mathcal U}}
	\newcommand{\PP}{\mathbb{P}}
	\newcommand{\ZZ}{\mathbb{Z}}
	\newcommand{\EE}{\mathbb{E}}
	\renewcommand{\d}{\mathrm{d}}
	\newcommand{\dm}{d_{\rm min}}
	\newcommand{\rhojunction}{\rho_{\rm j}}
	\newcommand{\rhojunctionLim}{\rho_{{\rm j},0}}
	\newcommand{\rhobranch}{\rho_{\rm b}}
	\newcommand{\rhobranchLim}{\rho_{{\rm b},0}}
	\newcommand{\rhononbridge}{\rho_{\rm n}}
	\newcommand{\rhononbridgeLim}{\rho_{{\rm n},0}}
	\newcommand{\percolationCluster}{C}
	\newcommand{\leafFreeCluster}{C_{\rm \ell f}}
	\newcommand{\bridgeFreeCluster}{C_{\rm bf}}
	\newcommand{\dfi}{D_\textsc{f1}}
	\newcommand{\dfp}{D_\textsc{f2}}
	\newcommand{\Df}{D_\textsc{F}}
	\newcommand{\dffi}{\overline{D}_\textsc{f1}}
	\newcommand{\dffp}{\overline{D}_\textsc{f2}}
	\newcommand{\yt}{y_{\rm t}}
	\newcommand{\yh}{y_{\rm h}}
	\newcommand{\dfprime}{d'_{\rm f}}
	\newcommand{\yhhat}{\hat{y}_{\rm h}}
	\newcommand{\ythat}{\hat{y}_{\rm t}}
	\newcommand{\yhstar}{y^*_{\rm h}}
	\newcommand{\ytstar}{y^*_{\rm t}}
	\newcommand{\etaQ}{\eta_{\rm Q}}
	\newcommand{\zc}{z_{\rm c}}
	\newcommand{\dc}{d_{c}}
	\newcommand{\dcp}{d_{p}}
	\newcommand{\ytP}{y^{\rm P}_{\rm t}}
	\newcommand{\yhP}{y^{\rm P}_{\rm h}}
	\newcommand{\dcI}{d^{\rm I}_{\rm c}}
	\newcommand{\bfx}{{\bf x}}
	\newcommand{\bfxbar}{\bar{\bf x}}
	\newcommand{\bfO}{{\bf o}}
	\newcommand{\bfo}{{\bf o}}
	\newcommand{\bfr}{\bf r}
	\newcommand{\origin}{\bf 0}
	\newcommand{\bfe}{\bf e}
	\newcommand{\bfk}{{\bf k}}
	\newcommand{\bfy}{\bf y}
	\newcommand{\bfu}{\bf u}
	\newcommand{\bmomega}{{\bm \omega}}
	\newcommand{\bfU}{{\bf u}}
	\newcommand{\ind}{\mathbbm{1}}
	\newcommand{\xiu}{\xi_{\rm u}}
	
	\newcommand{\DLone}{D_\textsc{L1}}
	\newcommand{\DLtwo}{D_\textsc{L2}}
	\newcommand{\DFone}{D_\textsc{F1}}
	\newcommand{\DFtwo}{D_\textsc{F2}}
	
	\newcommand{\Dvf}{D_{\textsc{v1}}}
	\newcommand{\Dff}{D_{\textsc{f1}}}
	\newcommand{\Dfs}{D_\textsc{f2}}
	
	\newcommand{\Dlf}{D_{\textsc{l1}}}
	\newcommand{\Dls}{D_{\textsc{l2}}}

	
	\title{Geometric Upper Critical Dimensions of the Ising Model}
	\date{\today}
	\author{Sheng Fang}
	\affiliation{Hefei National Research Center for Physical Sciences at the Microscales, University of Science and Technology of China, Hefei, Anhui 230026, China}
	\affiliation{MinJiang Collaborative Center for Theoretical Physics,
		College of Physics and Electronic Information Engineering, Minjiang University, Fuzhou 350108, China}
	\author{Zongzheng Zhou}
	\email{eric.zhou@monash.edu}
	\affiliation{ARC Centre of Excellence for Mathematical and Statistical Frontiers (ACEMS),
		School of Mathematics, Monash University, Clayton, Victoria 3800, Australia}
	\author{Youjin Deng}
	\email{yjdeng@ustc.edu.cn}
	\affiliation{Hefei National Research Center for Physical Sciences at the Microscales, University of Science and Technology of China, Hefei, Anhui 230026, China}	
	\affiliation{MinJiang Collaborative Center for Theoretical Physics,
		College of Physics and Electronic Information Engineering, Minjiang University, Fuzhou 350108, China}
	\affiliation{
		Shanghai Research Center for Quantum Sciences, Shanghai 201315, China}
\begin{abstract}
The upper critical dimension of the Ising model is known to be $d_c=4$, 
above which critical behavior is regarded as trivial. 
We hereby argue from extensive simulations that, in the random-cluster representation, 
the Ising model simultaneously exhibits two upper critical dimensions at $(d_c= 4, d_p=6)$, 
and critical clusters for $d \geq d_p$, except the largest one, 
are governed by exponents from percolation universality. 
We predict a rich variety of geometric properties and 
then provide strong evidence in 
dimensions from 4 to 7 and on complete graphs. 
Our findings significantly advance the understanding of 
the Ising model, which is
a fundamental system in many branches of physics. 

\end{abstract}


    
\maketitle

\maketitle

{\it Introduction--}The Ising model~\cite{friedli2017statistical} is one of the most 
fundamental models in statistical physics and condensed matter, 
and has great influence in almost every branch of modern physics. 
When introduced in 1925, the Ising model was 
shown to have no finite-temperature phase transition in one dimension (1D)~\cite{ising1925beitrag}. 
In 1944, a milestone was achieved by Onsager~\cite{onsager1944crystal}~\footnote{Onsager first announced the result in a conference in 1942, while the official paper was not published until 1944}  who obtained the exact free energy density on the square lattice without external field.
In 1952, Yang derived that, close to the critical temperature, 
the spontaneous magnetization vanishes as a power law 
with exponent $\beta = 1/8$~\cite{yang1952spontaneous}. 
In 1970s, the renormalization group (RG) theory was established, which is now the foundation of 
the modern theory of critical phenomena~\cite{wilson1971renormalization,wilson1971renormalizationa,wilson1972critical,wilson1975renormalization}. 
An important result of the RG theory is that 
the Ising model has the upper critical dimension $\dc = 4$, 
above which its critical behavior is controlled by the Gaussian fixed point (GFP). 
In 3D, extensive numerical studies have been available~\cite{deng2003simultaneous,ferrenberg2018pushing,hou2019geometric}, 
and, recently, the  conformal bootstrap program has led to new insights and an unprecedented precision
of critical exponents~\cite{kos2016precision,poland2019conformal}.

Percolation~\cite{broadbent1957percolation,stauffer2018introduction} has been 
intensively studied since 1950s due to its richness in both physics and mathematics. 
In bond percolation, the edges of a lattice are occupied with probability $p$, or vacant.
Two sites are connected if there is a path of occupied bonds from one to the other.
A maximal set of connected sites is called a cluster. The upper critical dimension of percolation 
is known to be $d_p = 6$~\cite{chayes1987upper}, and, for $d\geq 11$~\cite{AizenmanNewman1984,HaraSlade1990},
many rigorous results have been obtained. Besides their fundamental roles, 
both percolation and the Ising model are widely 
applied to many fields, including material science, neuroscience, 
complex network, epidemiology, ecology and biology 
etc.~\cite{Herega2015,Hopfield1982,SergeyStanley2010,Mello2021,ZengStanley2020,Ma2019,BrunkTwarock2021,ZhangStanley2019}.

In 1972, Fortuin and Kasteleyn (FK) derived 
the so-called random-cluster (RC) representation~\cite{Grimmett2006}
for the $Q$-state Potts model~\cite{Wu1982},
in which the Ising and percolation models are simply 
the special cases for $Q=2$ and $Q \! \rightarrow \! 1$, respectively.
This geometric representation has led to many exact results in 2D and efficient simulation 
algorithms~\cite{swendsen1987nonuniversal,wolff1989collective,chayes1998graphicala,zhang2020loop}. 
A natural question arises: what is the upper critical dimension $d_u$ of the general RC model. 
In 1970s, the RG analysis suggested that, depending on the inclusion of the $\phi^3$ term or not, $d_u$ could be 6 or 4 for general $Q$.
Special attention was paid to the FK-Ising model.
On the Bethe lattice and complete graph (CG)~\footnote{A complete graph with $V$ vertices is a graph in which each vertex is connected to all others.}, both of which can be regarded as the $d \rightarrow \infty$ limit,
$d_u=6$ was conjectured from a hyperscaling relation~\cite{chayes1999meanfield}.
On CGs, a percolation-like scaling window was rigorously shown~\cite{LuczakLuczak2008},
and a recent study revealed that, at criticality, 
medium-size clusters are CG-percolation-like~\cite{FangZhouDeng2021}.
In 5D, interesting two-scale scaling behaviors were observed~\cite{FangGrimmZhouDeng2020}.


\begin{figure*}[!htbp]
	\centering
	\includegraphics[width=\textwidth]{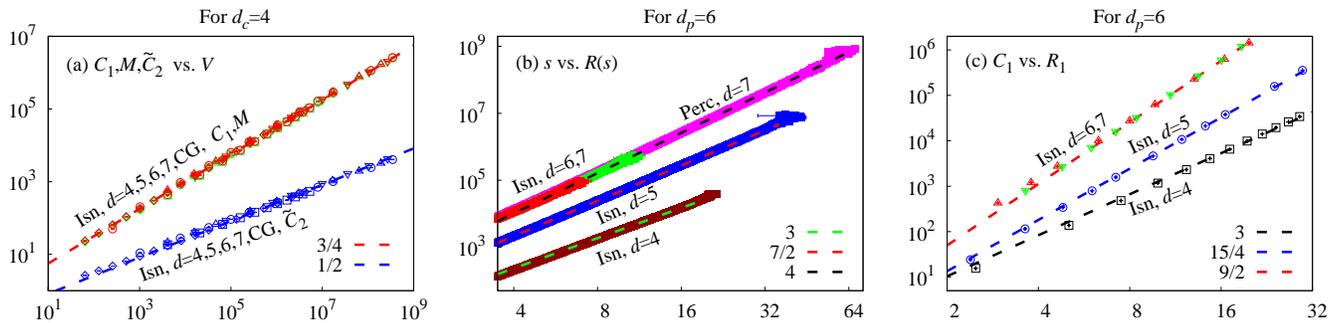}
	\caption{Evidence for the two upper critical dimensions, with power-law scaling illustrated by approximately straight lines in the log-log scale.
		(a), evidence for $d_c=4$ from finite-size scaling. 
		The largest-cluster size $C_1$, the rescaled second-largest-cluster size $\tilde{C}_2$
		and the magnetization $M$ are plotted versus system volumn $V$. 
		Up to non-universal constants, data of $C_1$ and $M$ collapse well onto 
		a line with slope $3/4$ for $d=4,5,6,7$ and on CGs,
		and data of $\tilde{C}_2$ collapse onto a line of slope $1/2$. 
		(b), evidence for $d_p=6$ from geometric fractals.  
		Size $s$ of medium clusters is shown versus gyration radius $R$, and the fractal dimension 
		$ \Dfs$ is $1+d/2$ for $4 \leq d < 6$ and $4$ for $d \geq 6$ (percolation universality).
		(c), evidence for $d_p=6$ from the largest cluster, 
		which has fractal dimension $\Dff=3d/4$ for  $4 \leq d < 6$ and  $\Dff=9/2$ for  $d \geq 6$.}
	\label{fig:main_evidence}
	\vspace{-0.30cm}
\end{figure*}   

In this Letter, we carry out extensive simulations 
for the Ising model on periodic hypercubic lattice of linear size $L$ 
in dimensions from 4 to 7 and on CGs, as well as for bond percolation in 7D.
The simulation is up to be of more than $10^8$ lattice sites. 
We observe a surprisingly rich variety of 
finite-size and thermodynamic critical behaviors,
and, on this basis, argue that, in the RC representation, 
the Ising model simultaneously exhibits two upper 
critical dimensions at $d_c=4$ and $d_p=6$, respectively.

On the one hand, as long as for $d \geq 4$, 
a bunch of geometric quantities
display finite-size scaling (FSS) behavior governed by a uniform set of mean-field exponents. 
For instance, the $L$-dependent scaling exponent for the largest cluster is $\Dlf=3d/4$ from the CG-Ising asymptotics, 
while it is $\Dls=1+d/2$ for the second-largest cluster from the GFP in the RG framework. 
On the other hand, critical clusters exhibit different geometric structures 
for $4 \leq d < 6$ and $d \geq 6$.
Consider the thermodynamic fractal dimension $\Df$, as defined from the asymptotic 
power-law dependence of the size of a cluster on its gyration radius.
For $ 4 \leq d < 6$, one has $\Dff=3d/4$ for the largest cluster
and $\Dfs=1+d/2$ for all the remaining ones.
However, for $d \geq 6$, one has $\Dff =9/2$ and $\Dfs=4$;
the latter is from high-$d$ percolation universality.
This is summarized in Table~\ref{tab:introduction}.
A variety of other critical behaviors are observed. 
For instance, for $ d \geq  6$, the number of spanning clusters 
and the winding number of the largest cluster 
are both divergent as $L$ increases,
while they are of ${\cal O}(1)$ in lower dimensions.
Further, there exist two scaling windows:
the leading one is of CG-Ising type,
and the other is of Gaussian type for $4 \leq d < 6$ and 
of percolation type for $d \geq 6$,
align with rigorous result for CGs.

\begin{table}[t]  
	\centering
	\begin{tabular}{l|rlrl}
		\hline 
		& \multicolumn{2}{c}{$4\le d < 6$\;\;\;\;}  &                 & $d \geq 6$  \\
		\hline 
		$\Dlf$ & $3d/4$        &         & \;\; $\Leftarrow$\;\; & same     \\
		\vspace{1mm}
		$\Dff$ & $3d/4$        & (CG-Ising asy.)      &              & $9/2$    \\ 
		$\Dls$ & $1\!+\!d/2$   &                  & $\Leftarrow$\;\; &  same     \\   
		$\Dfs$ & $1\!+\!d/2$   & (Gaussian f.p.$\:$)  &              & $4$\; (percolation) \\ 
		\hline 
	\end{tabular}
	\caption{Conjectured exact fractal dimensions for $4\leq d <6$ and $d \geq 6$,
		as inspired by CG-Ising asymptotics, Gaussian fixed point and results for high-$d$ percolation.}
	\label{tab:introduction}
	\vspace{-0.5cm}
\end{table}

{\it Models--} The Hamiltonian of the Ising model reads
\begin{equation}
	{\cal H} = -K\sum_{i\sim j}S_iS_j \;, \hspace{5mm} (S_i= \pm 1)
\end{equation}
where $K \! > \! 0$ represents the ferromagnetic coupling and the summation is over all neighboring pairs.
By the FK transformation, it can be mapped onto the $Q$-state RC model with $Q=2$,
with the partition function
\begin{equation}
	{\cal Z} = \sum_{A \subseteq G} p^{|A|} 
	(1-p)^{|E\backslash A|} Q^{c(A)}\;,
\end{equation}
where the lattice is denoted as $G \equiv (V,E)$, the summation is over all spanning subgraphs $A \subseteq G$, 
$|A|$ and $c(A)$ respectively represent the number of occupied bonds and of clusters,
and the bond probability is $p = 1- e^{-2K}$.

We simulate the FK-Ising model on $d$-dimensional tori with $4\leq d\leq 7$ 
and on CGs, at and near the critical points as in Refs.~\cite{Lv2019Two,lundow2015discontinuity}.
A combination of the Wolff and Swendsen-Wang algorithms~\cite{swendsen1987nonuniversal,wolff1989collective} 
is applied, and the latter is mainly used to generate FK-bond configurations.
The maximum system volume is $V=48^4,51^5,24^6,16^7,2^{22}$ for $d=4,5,6,7$ and CGs, respectively. 
We also simulate bond percolation in 7D at criticality~\cite{MertensMoore2018}.

We sample the number $n(s,V)$ of clusters of size $s$ per site, 
and the sizes of the largest- and the second-largest clusters as
$C_1 = \langle {\cal C}_1 \rangle$ and $C_2$, respectively, 
with $\langle \cdot \rangle$ for ensemble average. 
Further, to study geometric fractal structures,
we use the breadth-first search method to grow FK clusters and
measure their gyration radius in an \emph{unwrapped} way,
effectively taking into account periodic boundary effects \cite{FangGrimmZhouDeng2020}. 
Given a cluster ${\cal C}$, we randomly choose a seed site 
and assign it a $d$-dimensional {\it zero} coordinate (${\bfx \equiv 0}$),  
and each newly included site $v$ is assigned an unwrapped coordinate as
$\bfx_v = \bfx_u + \bfe_i \, (-{\bfe_i})$, if $v$ is grown from $u$ along (against) 
the $i$th direction, with $\bfe_i$ the corresponding unit vector.
The unwrapped gyration radius is calculated as
$\scrR \equiv \sqrt{ \langle |\bfx_u|^2 \rangle - \langle |\bfx_u| \rangle^2} $,
where the average is over all sites in cluster ${\cal C}$.
We also measure the unwrapped expansion distance $\mathcal{U}$ along the first-coordinate direction for each cluster.

{\it Evidence for $d_c=4$ from finite-size scaling--} In the spin representation, 
$d_c=4$ is widely known for the Ising model. 
Nevertheless, finite-size scaling behavior for $d \geq 4$ 
has been a long-standing debate \cite{wittmann2014finitesize,flores-sola2016role,grimm2017geometric,zhou2018randomlength,FangGrimmZhouDeng2020}. 
It is now believed~\cite{FangGrimmZhouDeng2020,Lv2019Two,FangDengZhou2021} 
that the critical free energy on high-$d$ tori contain two scaling terms,
having RG exponents $(y_t=2, y_h=1 \! +\!d/2)$ from the GFP
and $(y^*_t=d/2, y^*_h=3d/4)$ from the CG-Ising asymptotics.
An important consequence is that the critical two-point function 
behaves as $G(\bfx, L) \approx \|\bfx\|^{2-d} + L^{-d/2}$,
algebraically decaying with distance $\|\bfx\|$, 
with exponent ${2-d}$ from GFP, and then saturating 
to a plateau of height $L^{-d/2}$ from CG-Ising asymptotics \cite{Papathanakos2006,GrimmElciZhouGaroniDeng2017,ZhouGrimmFangDengGaroni2018}. This implies that the magnetic susceptibility, which is exactly the average cluster size in the FK representation, scales as $L^{d/2}$.

Figure~\ref{fig:main_evidence}(a) shows the critical magnetization 
$M \equiv \langle | \sum_i  S_i | \rangle $ versus volume $V$.
The good data collapsing for $d=4,5,6,7$ and on CGs, displaying $M \sim V^{3/4}$.
Moreover, the $C_1$ data collapse well onto those for $M$. 
This confirms the conventional upper dimension $d_c=4$,
and demonstrates the uniform scaling $\sim V^{3/4}$ for $d \geq d_c$,
which can be proved for CGs~\cite{BollobasGrimmettJanson1996,LuczakLuczak2008}.


From extensive simulations and results in Ref.~\cite{FangGrimmZhouDeng2020}, 
we conjecture that, for $d > 4$,  the FSS of $C_2$ behaves as 
$C_2 \sim  L^{1+d/2} = \sqrt{V} V^{1/d}$, corresponding to the GFP. This is seemingly consistent with the scaling $C_2 \sim \sqrt{V} \ln V$ for CGs~\cite{LuczakLuczak2008}, where $\ln V$  might relate to the term $V^{1/d}$ for finite-$d$. Rescaled quantities are then defined as $\tilde{C}_2 \equiv C_2/L$
for $d \geq 4$  and $\tilde{C}_2 \equiv C_2/\ln V$ for CGs.
Indeed, the $\tilde{C}_2$ data for $d=4,5,6,7$ and on CGs
collapse well on a line with slope $1/2$, shown in Fig.~\ref{fig:main_evidence}(a).

\begin{figure}
	\centering
	\includegraphics[width=0.47\textwidth]{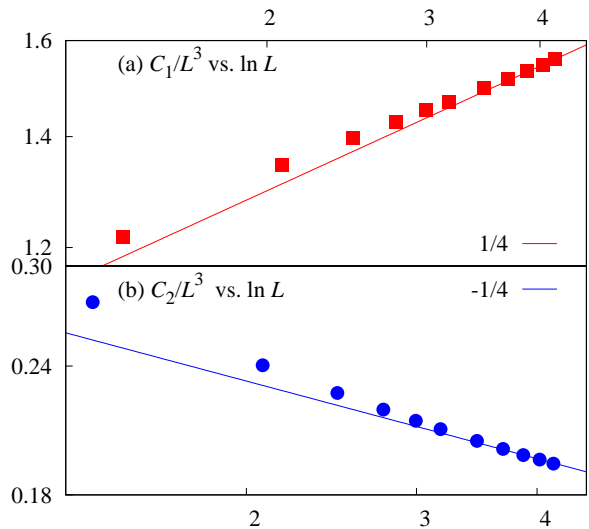}
	\caption{Log-log plot of the rescaled sizes of the largest and second largest clusters at $d=4$ versus $\ln L$.}
	\label{fig:C12d4}
\end{figure}

At the upper critical dimensions, logarithmic corrections are usually expected. For the Ising model in the spin representation, field theory predicts the form of logarithmic corrections for many quantities at $d_c = 4$~\cite{kenna2013universal,kenna2004finite}, such as the magnetization $M \sim L^3 (\ln L)^{1/4}$, and the susceptibility $\chi \sim L^{2}(\ln L)^{1/2}$. We now examine the effect of logarithmic corrections to $C_1$ and $C_2$. In Fig.~\ref{fig:C12d4}, we plot in log-log scale $C_1$ and $C_2$, rescaled by their expected power-law scaling, versus $\ln L$. Our data suggest that $C_1 \sim L^3 (\ln L)^{1/4}$, consistent with the field-theory prediction for $M$, and $C_2 \sim L^{3} (\ln L)^{-1/4}$ which has no direct counterpart in the spin representation.

{\it Evidence for $d_p=6$ from geometric fractals--} In comparison with FSS,
intrinsic geometric properties of clusters 
are better characterized by the power-law dependence of cluster size on 
gyration radius as $ s \sim R^{D_\textsc{F}} $,
which is shown in Fig.~\ref{fig:main_evidence}(b) for 
medium-size clusters--i.e., clusters with size $1\ll s \ll C_1$.
Distinct fractal structures are revealed: the fractal dimension $\Dfs$ is $1+d/2$ for $4\leq d<6$, 
and becomes constant $4$ for $d \geq 6$.
While the former is from the GFP,
the latter is consistent with percolation universality \cite{aharony1984scaling}, 
as well illustrated by the 7D-percolation data in Fig.~\ref{fig:main_evidence}(b).

Actually, the largest cluster also has different fractal dimensions below and above $d_p=6$. 
The plot of the $C_1$ data against the gyration radius $R_1$ in Fig.~\ref{fig:main_evidence}(c)
gives $\Dff=3d/4$ for $4\leq d<6$ and $9/2$ for $d \geq 6$, 
with $9/2$ calculated from $3d/4$ with $d=6$.

Therefore, we conclude that $d_p=6$ is also an upper critical dimension for the FK-Ising model.

\begin{figure}
	\centering
	\includegraphics[width=0.50\textwidth]{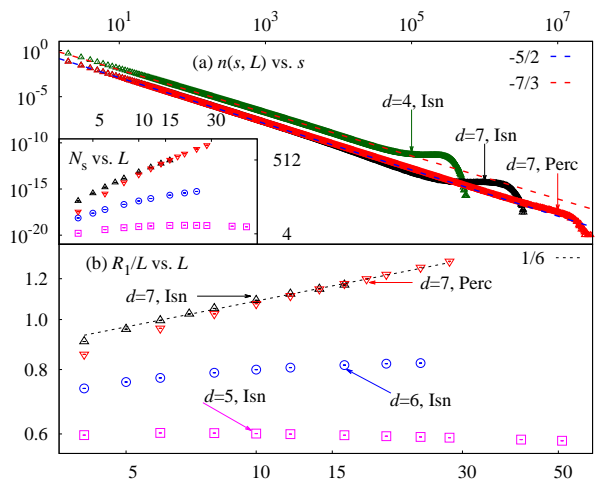}
	\caption{Evidence for $d_p=6$ from topological properties.
		(a), cluster-number density $n(s,L)$ versus $s$, where the Fisher exponent $\tau$ 
		is clearly different for $d=4$ and 7. 
		The inset is for the number $N_s$ of spanning clusters.
		(b) Winding number of the largest cluster as represented by $R_1/L$.
		Both the winding number and the spanning-cluster number are of size
		${\cal O}(1)$ for $d <6$ but diverge for $d>6$. 
	}
	\label{fig:R1Ns2}
\end{figure} 

{\it Evidence for $d_p=6$ from topological properties--}
The essential assumption of the standard FSS theory is that 
the divergent correlation length--e.g., as characterized by $R_1$--is
cut off as ${\cal O}(L)$, resulting in that the number of percolating clusters is of ${\cal O}(1)$.
This has been widely used as a powerful tool in numerical study of critical phenomena. 

We first look at the cluster-number density $n(s,L) \sim s^{-\tau} 
\tilde{n} (s/L^{\Dlf})$, where $\tau$ is the Fisher exponent, 
$\Dlf$ is the finite-size fractal dimension and $\tilde{n}$ is a universal function.
The hyperscaling relation, $\tau = 1+d/\Dlf$, is further believed to hold, 
giving $\tau=7/3$ for $d \geq 4$.
As shown in Fig.~\ref{fig:R1Ns2}(a),
while being indeed true for 4D, the hyperscaling relation is broken for 7D, which has $\tau\approx 5/2$.
From the data collapsing for the FK-Ising and percolation models in 7D,
it can be restored by using $D_\textsc{lp}=2d/3$ for percolation universality.

To illustrate the emergence of clusters with nontrivial topology,
we measure the number $N_s$ of spanning clusters, of which the unwrapped expansion distance $\mathcal{U} \ge L$.
The inset of Fig.~\ref{fig:R1Ns2}(a) shows that, while $N_s = {\cal O}(1)$ for $d <6$, 
it diverges as $L$ increases for $d>6$. 
From the scaling $s \sim R^4$ in Fig.~1(b), 
it is suggested that the typical size of spanning clusters must be $ s > L^4$,
and, thus, $N_s$ can be calculated as $L^d \int_{L^4} n(s,L) \, {\rm d} s$.
With $\tau=5/2$ for $d > 6$, this gives $N_s \sim L^{d-6}$, consistent with the inset of Fig.~\ref{fig:R1Ns2}(a).

Topological properties can be further illustrated by the winding number, 
as characterized by ratio $R_1/L$. As shown in Fig.~\ref{fig:R1Ns2}(b), 
one has $R_1/L = {\cal O}(1)$ for $d <6$, consistent with the observation 
of $\Dlf=\Dff$--i.e., the finite-size and thermodynamic fractal dimensions are identical.
For $d > 6$, however, one has $C_1 \sim L^{3d/4} \sim R_1^{9/2}$ (Table~\ref{tab:introduction}), 
and thus expects $R_1 \sim L^{d/6}$. 
In other words, as $L$ increases, the largest cluster winds around the tori for more and more times.
This is well confirmed in Fig.~\ref{fig:R1Ns2}(b). 

\begin{figure}[t]
	\centering
	\includegraphics[width=0.5\textwidth]{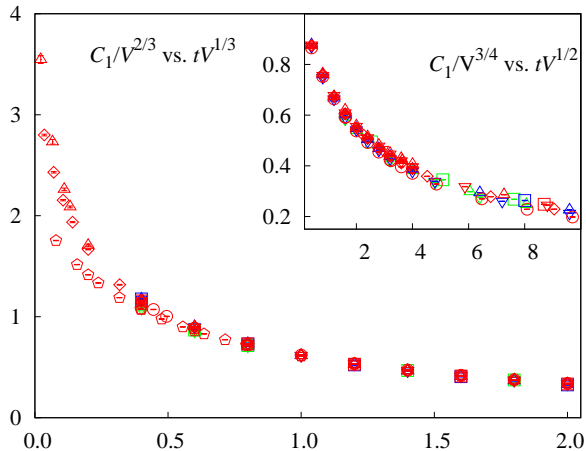}  
	\caption{Percolation scaling window for $d \geq 6$. 
		Within window $t \sim {\cal O}(V^{-1/3})$, 
		the largest cluster scales percolation-like as $C_1 \sim V^{2/3}$.
		The data points of various shapes are for different system sizes $V$, 
		and the colors are for 6D (blue), 7D (red) and CGs (green).
		The inset demonstrates the CG-Ising scaling window of $ {\cal O}(V^{-1/2})$, 
		in which $C_1(t,V) \sim V^{3/4} \tilde{C}_1 (tV^{1/2})$.}
	\label{fig:C1sw}
\end{figure}

{\it Percolation-like scaling for the largest cluster--} 
The above scaling behaviors  at criticality also hold within a scaling window of 
size ${\cal O}(1/L^{y^*_t})$, with $y^*_t=d/2$ from the CG-Ising asymptotics. 
As an example, the inset of Fig.~\ref{fig:C1sw} shows the scaling 
$C_1(t,V) \sim V^{3/4} \tilde{C}_1 (tV^{1/2})$ for the largest cluster of the Ising model with $d=6,7$ and CG,
where $t \equiv (K_c-K)/K_c$ and $t \geq 0$ is for the high-temperature phase. This also has been observed for 5D FK Ising model in Ref.~\cite{FangGrimmZhouDeng2020}.

Actually, in each  dimension $d \geq 4$, there also exists another scaling window, which is less sharp and thus can survive slightly further away from $K_c$. For $4 \leq d < 6$, it is of size ${\cal O}(1/L^{y_t})$, with $y_t=2$ from the GFP, where all the clusters, including the largest one, would scale as $s \sim L^{y_h} \sim R^{y_h}$ with $y_h=1+d/2$.

For $d>6$, the second scaling window is of size ${\cal O}(1/L^{d/3})$, with exponent $d/3$ from high-$d$ percolation, where all the clusters, including the largest one, are expected to be percolation-like as $s \sim R^4 \sim L^{2d/3}$. This is illustrated by Fig.~\ref{fig:C1sw}, displaying $C_1(t,V) \sim V^{2/3} \tilde{C}_1 (tV^{1/3})$. Note that, unlike the CG-Ising scaling window, the percolation scaling window only occurs at the high temperature side. The scattering for small values of $tV^{1/3}$ is due to the CG-Ising scaling window. On CGs, the existence of the percolation scaling window has been rigorously proved~\cite{LuczakLuczak2008} and an RG-like argument has been provided~\cite{FangZhouDeng2021}.
	
The second-largest cluster scales as $C_2 \sim L^{1+d/2}$ in the CG-Ising scaling window (including $K_c$). It is therefore expected for $d>6$ that, the maximum of $C_2$ would occur in the percolation scaling window and diverge as $\sim L^{2d/3}$.

Other interesting phenomena emerge. For instance, as criticality is approached from the low-temperature side--i.e., $t \rightarrow 0^{-}$, one can expect that the second-largest cluster scales  as $C_2(t,V) \sim L^{1+d/2} \tilde{C}_2(tL^{d/2})$. Suppose the relation $C_2 \sim R_2^{4}$ holds within the scaling window, then we have $R_2 \sim L^{(d+2)/8} \tilde{R}_2(tL^{d/2}) $ for $d\ge6$. To recover the thermodynamic critical behavior, one expects $\tilde{R}_2(x) \sim x^{-(d+2)/4d}$, such that $R_2 \sim |t|^{-\nu_2'}$ with $\nu_2'=(d+2)/4d$. Thus $\nu'_2 = 1/3$ for $d=6$ and converges to $1/4$ 
as $d \rightarrow \infty$. The exponent $1/4$ was also obtained on the Bethe lattice with fixed boundary conditions~\cite{chayes1999meanfield}. In addition, it was observed on CGs~\cite{FangZhouDeng2021} that a tiny sector emerges in the whole configuration space and slowly vanishes as $L$ increases. Conditioned on being in this sector, quantities are observed to exhibit CG-percolation behaviour. On lattices, our preliminary simulations suggest that there exists a tiny sector for all $d\geq 4$, which is of Gaussian and percolation types for $4 \leq d < 6$ and $d \geq 6$, respectively.


{\it Conclusion--} Based on a combination of extensive simulations 
from $d=4$ to 7 and insights from RG theory,
and rigorous and numerical results for CG,
we propose that, in the FK random-cluster representation, 
the Ising model simultaneously has 
two upper critical dimensions at $(d_c=4, d_p=6)$.
Besides being an answer for the long-standing debate, dated back to 1970s, this picture provides 
a counter-intuitive and advanced understanding 
for the Ising model, which is probably the most fundamental
system in statistical and condensed-matter physics.
Note that the scenario of two upper critical dimensions was also proposed in the field-theoretical treatment of the $\rm{CP}^{1}$ model \cite{nahum2013phase}.

In FK-Ising clusters for $d \geq 4$, the thermodynamic and finite-size scaling behaviors are surprisingly rich,
partially summarized in Tables~\ref{tab:introduction} and~\ref{tab:summary}.
Two pronounced features can be seen: 1), as long as $d \geq 4$, there exist
two-scale properties, two scaling windows and two configuration sectors;
and 2), for $d>6$, the scaling behaviors of all clusters, except the largest one, 
are in percolation universality, unexpected from the first sight. 
Interestingly, while the geometric properties are very sophisticated, 
critical behaviors in the spin representation are much simpler:
no percolation-like behaviors exist and the upper critical dimension $d_p=6$ cannot be seen.

\begin{table}[t]
	\centering
	\begin{tabular}{c|c|c}
		\hline 
		& \multicolumn{1}{c}{ $4\le d < 6$}   & $d \geq 6$  \\
		\hline 
		$\tau-1$  & \multicolumn{1}{c}{$d/(1+d/2)$}     & $3/2$    \\
		$R_1$     & \multicolumn{1}{c}{$\sim L$}        & $\sim L^{d/6}$    \\
		$N_{s}$   & \multicolumn{1}{c}{${\cal O}(1)$}   & $\sim L^{d-6}$    \\  
		scaling windows   &  \multicolumn{1}{c}{ CG-Isn.+ GFP} \;\;   
		&  \;\; CG-Isn. + Perc.                \\ 
		\hline 
	\end{tabular}
	\caption{Some scaling behaviors for $4 \leq d <6$ and $d \geq 6$,
		including Fisher exponent $\tau$, finite-size scaling of the gyration radius $R_1$ and 
		the number $N_s$ of spanning clusters, and two scaling windows.}
	\label{tab:summary}
	\vspace{-0.5cm}
\end{table}

Several open questions arise. 
First, what are the precise forms of logarithmic corrections 
in critical FK clusters at $d_c=4$ and $d_p=6$? In this work, we study the logarithmic corrections of sizes of the largest two clusters, but it will be interesting to carry out a systematic study of the effect of logarithmic corrections to various geometric quantities, especially at $d_p = 6$.
Second, in the loop representation of the Ising model, 
which is another geometric representation 
and can be coupled to the RC model via the loop-cluster joint model~\cite{zhang2020loop},
what would be geometric effects for $d \geq 4$?
Finally, most of the exact exponents in Tables~\ref{tab:introduction} and~\ref{tab:summary}
are conjectured and rigorous proofs remain elusive.

\paragraph{Acknowledgements}
This work has been supported by the National Natural Science Foundation of China 
(under Grant No. 11625522), the Science and Technology Committee of Shanghai 
(under grant No. 20DZ2210100), the National Key R\&D Program of China (under Grant No. 2018YFA0306501).
We thank Eren M. El\c{c}i, Jens Grimm, Timothy Garoni, 
Martin Weigel and Jonathan Machta for valuable discussions, in particular for Jesper Jacobsen. 
When finalizing the collection and analysis of our Monte Carlo data, we learned from private communications 
that Jesper Jacobsen and Kay Wiese (ENS, Paris) are working on the same topic 
using a field-theoretical approach and propose another scenario—i.e., 
for $ 4 \leq d < 6$, the scaling behavior of some geometric observables 
could be described by non-trivial critical exponents other than those from the CG-Ising asymptotic 
and from the GFP. Taking into account the logarithmic correction for the scaling of 
the second-largest cluster, which is conjectured solely based on simulations, 
this interesting scenario cannot be ruled out. \par

We dedicate this work to Professor Henk W.J. Bl\"ote, who passed away  on June 10, 2022. 
Bl\"ote was internationally renowned for his numerous contributions to statistical mechanics, 
holding official positions at Delft University of Technology and Leiden University 
until his retirement in 2008, as well as a lifetime service to physics.
Since his first paper on the specific heat singularities of Ising antiferromagnets in 1967, 
Bl\"ote has maintained a particular passion for the Ising model among his research interests 
in different physical topics. 
As this work demonstrates, he has successfully conveyed his spirit to his students (Y.D) 
and his second-generation students (S.F and Z.Z). 
Bl\"ote has maintained a very close relationship with China over the past few decades, even learning to speak the Chinese language. 
Bl\"ote gave enormous guidance to the students and researchers he supervised, treating them as his children. 
His research fellows, especially his two Chinese PhD students (Youjin Deng and Xiaofeng Qian) 
and his Chinese postdoc Wen'an Guo are so grateful for having had Bl\"ote as their supervisor. 
Bl\"ote was very generous, kind, and always ready to provide us with support and love. 
Bl\"ote was our physics mentor and remains our lifetime mentor. 
The seed of physics he sowed in China has grown into academic trees of several generations; 
the seed of love he planted  in China has grown into a sea of sunflowers 
that warms the hearts of countless people.

\bibliographystyle{apsrev4-1}

\begin{thebibliography}{54}%
	\makeatletter
	\providecommand \@ifxundefined [1]{%
		\@ifx{#1\undefined}
	}%
	\providecommand \@ifnum [1]{%
		\ifnum #1\expandafter \@firstoftwo
		\else \expandafter \@secondoftwo
		\fi
	}%
	\providecommand \@ifx [1]{%
		\ifx #1\expandafter \@firstoftwo
		\else \expandafter \@secondoftwo
		\fi
	}%
	\providecommand \natexlab [1]{#1}%
	\providecommand \enquote  [1]{``#1''}%
	\providecommand \bibnamefont  [1]{#1}%
	\providecommand \bibfnamefont [1]{#1}%
	\providecommand \citenamefont [1]{#1}%
	\providecommand \href@noop [0]{\@secondoftwo}%
	\providecommand \href [0]{\begingroup \@sanitize@url \@href}%
	\providecommand \@href[1]{\@@startlink{#1}\@@href}%
	\providecommand \@@href[1]{\endgroup#1\@@endlink}%
	\providecommand \@sanitize@url [0]{\catcode `\\12\catcode `\$12\catcode
		`\&12\catcode `\#12\catcode `\^12\catcode `\_12\catcode `\%12\relax}%
	\providecommand \@@startlink[1]{}%
	\providecommand \@@endlink[0]{}%
	\providecommand \url  [0]{\begingroup\@sanitize@url \@url }%
	\providecommand \@url [1]{\endgroup\@href {#1}{\urlprefix }}%
	\providecommand \urlprefix  [0]{URL }%
	\providecommand \Eprint [0]{\href }%
	\providecommand \doibase [0]{http://dx.doi.org/}%
	\providecommand \selectlanguage [0]{\@gobble}%
	\providecommand \bibinfo  [0]{\@secondoftwo}%
	\providecommand \bibfield  [0]{\@secondoftwo}%
	\providecommand \translation [1]{[#1]}%
	\providecommand \BibitemOpen [0]{}%
	\providecommand \bibitemStop [0]{}%
	\providecommand \bibitemNoStop [0]{.\EOS\space}%
	\providecommand \EOS [0]{\spacefactor3000\relax}%
	\providecommand \BibitemShut  [1]{\csname bibitem#1\endcsname}%
	\let\auto@bib@innerbib\@empty
	\bibitem [{\citenamefont {Friedli}\ and\ \citenamefont
		{Velenik}(2017)}]{friedli2017statistical}%
	\BibitemOpen
	\bibfield  {author} {\bibinfo {author} {\bibfnamefont {S.}~\bibnamefont
			{Friedli}}\ and\ \bibinfo {author} {\bibfnamefont {Y.}~\bibnamefont
			{Velenik}},\ }\href@noop {} {\emph {\bibinfo {title} {Statistical Mechanics
				of Lattice Systems: A Concrete Mathematical Introduction}}}\ (\bibinfo
	{publisher} {Cambridge University Press},\ \bibinfo {year}
	{2017})\BibitemShut {NoStop}%
	\bibitem [{\citenamefont {Ising}(1925)}]{ising1925beitrag}%
	\BibitemOpen
	\bibfield  {author} {\bibinfo {author} {\bibfnamefont {E.}~\bibnamefont
			{Ising}},\ } {\bibfield  {journal}
		{\bibinfo  {journal} {Zeitschrift f\"ur Physik}\ }\textbf {\bibinfo {volume}
			{31}},\ \bibinfo {pages} {253} (\bibinfo {year} {1925})}\BibitemShut
	{NoStop}%
	\bibitem [{\citenamefont {Onsager}(1944)}]{onsager1944crystal}%
	\BibitemOpen
	\bibfield  {author} {\bibinfo {author} {\bibfnamefont {L.}~\bibnamefont
			{Onsager}},\ } {\bibfield  {journal}
		{\bibinfo  {journal} {Physical Review}\ }\textbf {\bibinfo {volume} {65}},\
		\bibinfo {pages} {117} (\bibinfo {year} {1944})}\BibitemShut {NoStop}%
	\bibitem [{Note1()}]{Note1}%
	\BibitemOpen
	\bibinfo {note} {Onsager first announced the result in a conference in 1942,
		while the official paper was not published until 1944}\BibitemShut {NoStop}%
	\bibitem [{\citenamefont {Yang}(1952)}]{yang1952spontaneous}%
	\BibitemOpen
	\bibfield  {author} {\bibinfo {author} {\bibfnamefont {C.~N.}\ \bibnamefont
			{Yang}},\ } {\bibfield  {journal}
		{\bibinfo  {journal} {Physical Review}\ }\textbf {\bibinfo {volume} {85}},\
		\bibinfo {pages} {808} (\bibinfo {year} {1952})}\BibitemShut {NoStop}%
	\bibitem [{\citenamefont
		{Wilson}(1971{\natexlab{a}})}]{wilson1971renormalization}%
	\BibitemOpen
	\bibfield  {author} {\bibinfo {author} {\bibfnamefont {K.~G.}\ \bibnamefont
			{Wilson}},\ } {\bibfield  {journal}
		{\bibinfo  {journal} {Physical Review B}\ }\textbf {\bibinfo {volume} {4}},\
		\bibinfo {pages} {3174} (\bibinfo {year} {1971}{\natexlab{a}})}\BibitemShut
	{NoStop}%
	\bibitem [{\citenamefont
		{Wilson}(1971{\natexlab{b}})}]{wilson1971renormalizationa}%
	\BibitemOpen
	\bibfield  {author} {\bibinfo {author} {\bibfnamefont {K.~G.}\ \bibnamefont
			{Wilson}},\ } {\bibfield  {journal}
		{\bibinfo  {journal} {Physical Review B}\ }\textbf {\bibinfo {volume} {4}},\
		\bibinfo {pages} {3184} (\bibinfo {year} {1971}{\natexlab{b}})}\BibitemShut
	{NoStop}%
	\bibitem [{\citenamefont {Wilson}\ and\ \citenamefont
		{Fisher}(1972)}]{wilson1972critical}%
	\BibitemOpen
	\bibfield  {author} {\bibinfo {author} {\bibfnamefont {K.~G.}\ \bibnamefont
			{Wilson}}\ and\ \bibinfo {author} {\bibfnamefont {M.~E.}\ \bibnamefont
			{Fisher}},\ } {\bibfield
		{journal} {\bibinfo  {journal} {Physical Review Letters}\ }\textbf {\bibinfo
			{volume} {28}},\ \bibinfo {pages} {240} (\bibinfo {year} {1972})}\BibitemShut
	{NoStop}%
	\bibitem [{\citenamefont {Wilson}(1975)}]{wilson1975renormalization}%
	\BibitemOpen
	\bibfield  {author} {\bibinfo {author} {\bibfnamefont {K.~G.}\ \bibnamefont
			{Wilson}},\ } {\bibfield  {journal}
		{\bibinfo  {journal} {Reviews of Modern Physics}\ }\textbf {\bibinfo {volume}
			{47}},\ \bibinfo {pages} {773} (\bibinfo {year} {1975})}\BibitemShut
	{NoStop}%
	\bibitem [{\citenamefont {Deng}\ and\ \citenamefont
		{Bl{\"o}te}(2003)}]{deng2003simultaneous}%
	\BibitemOpen
	\bibfield  {author} {\bibinfo {author} {\bibfnamefont {Y.}~\bibnamefont
			{Deng}}\ and\ \bibinfo {author} {\bibfnamefont {H.~W.~J.}\ \bibnamefont
			{Bl{\"o}te}},\ } {\bibfield
		{journal} {\bibinfo  {journal} {Physical Review E}\ }\textbf {\bibinfo
			{volume} {68}},\ \bibinfo {pages} {036125} (\bibinfo {year}
		{2003})}\BibitemShut {NoStop}%
	\bibitem [{\citenamefont {Ferrenberg}\ \emph {et~al.}(2018)\citenamefont
		{Ferrenberg}, \citenamefont {Xu},\ and\ \citenamefont
		{Landau}}]{ferrenberg2018pushing}%
	\BibitemOpen
	\bibfield  {author} {\bibinfo {author} {\bibfnamefont {A.~M.}\ \bibnamefont
			{Ferrenberg}}, \bibinfo {author} {\bibfnamefont {J.}~\bibnamefont {Xu}}, \
		and\ \bibinfo {author} {\bibfnamefont {D.~P.}\ \bibnamefont {Landau}},\
	} {\bibfield  {journal} {\bibinfo
			{journal} {Physical Review E}\ }\textbf {\bibinfo {volume} {97}},\ \bibinfo
		{pages} {043301} (\bibinfo {year} {2018})}\BibitemShut {NoStop}%
	\bibitem [{\citenamefont {Hou}\ \emph {et~al.}(2019)\citenamefont {Hou},
		\citenamefont {Fang}, \citenamefont {Wang}, \citenamefont {Hu},\ and\
		\citenamefont {Deng}}]{hou2019geometric}%
	\BibitemOpen
	\bibfield  {author} {\bibinfo {author} {\bibfnamefont {P.}~\bibnamefont
			{Hou}}, \bibinfo {author} {\bibfnamefont {S.}~\bibnamefont {Fang}}, \bibinfo
		{author} {\bibfnamefont {J.}~\bibnamefont {Wang}}, \bibinfo {author}
		{\bibfnamefont {H.}~\bibnamefont {Hu}}, \ and\ \bibinfo {author}
		{\bibfnamefont {Y.}~\bibnamefont {Deng}},\ } {\bibfield  {journal} {\bibinfo  {journal}
			{Physical Review E}\ }\textbf {\bibinfo {volume} {99}},\ \bibinfo {pages}
		{042150} (\bibinfo {year} {2019})}\BibitemShut {NoStop}%
	\bibitem [{\citenamefont {Kos}\ \emph {et~al.}(2016)\citenamefont {Kos},
		\citenamefont {Poland}, \citenamefont {{Simmons-Duffin}},\ and\ \citenamefont
		{Vichi}}]{kos2016precision}%
	\BibitemOpen
	\bibfield  {author} {\bibinfo {author} {\bibfnamefont {F.}~\bibnamefont
			{Kos}}, \bibinfo {author} {\bibfnamefont {D.}~\bibnamefont {Poland}},
		\bibinfo {author} {\bibfnamefont {D.}~\bibnamefont {{Simmons-Duffin}}}, \
		and\ \bibinfo {author} {\bibfnamefont {A.}~\bibnamefont {Vichi}},\ } {\bibfield  {journal} {\bibinfo  {journal}
			{Journal of High Energy Physics}\ }\textbf {\bibinfo {volume} {2016}},\
		\bibinfo {pages} {36} (\bibinfo {year} {2016})}\BibitemShut {NoStop}%
	\bibitem [{\citenamefont {Poland}\ \emph {et~al.}(2019)\citenamefont {Poland},
		\citenamefont {Rychkov},\ and\ \citenamefont {Vichi}}]{poland2019conformal}%
	\BibitemOpen
	\bibfield  {author} {\bibinfo {author} {\bibfnamefont {D.}~\bibnamefont
			{Poland}}, \bibinfo {author} {\bibfnamefont {S.}~\bibnamefont {Rychkov}}, \
		and\ \bibinfo {author} {\bibfnamefont {A.}~\bibnamefont {Vichi}},\ } {\bibfield  {journal} {\bibinfo
			{journal} {Reviews of Modern Physics}\ }\textbf {\bibinfo {volume} {91}},\
		\bibinfo {pages} {015002} (\bibinfo {year} {2019})}\BibitemShut {NoStop}%
	\bibitem [{\citenamefont {Broadbent}\ and\ \citenamefont
		{Hammersley}(1957)}]{broadbent1957percolation}%
	\BibitemOpen
	\bibfield  {author} {\bibinfo {author} {\bibfnamefont {S.~R.}\ \bibnamefont
			{Broadbent}}\ and\ \bibinfo {author} {\bibfnamefont {J.~M.}\ \bibnamefont
			{Hammersley}},\ }in\ \href@noop {} {\emph {\bibinfo {booktitle} {Mathematical
				proceedings of the Cambridge philosophical society}}},\ Vol.~\bibinfo
	{volume} {53}\ (\bibinfo {organization} {Cambridge University Press},\
	\bibinfo {year} {1957})\ pp.\ \bibinfo {pages} {629--641}\BibitemShut
	{NoStop}%
	\bibitem [{\citenamefont {Stauffer}\ and\ \citenamefont
		{Aharony}(2018)}]{stauffer2018introduction}%
	\BibitemOpen
	\bibfield  {author} {\bibinfo {author} {\bibfnamefont {D.}~\bibnamefont
			{Stauffer}}\ and\ \bibinfo {author} {\bibfnamefont {A.}~\bibnamefont
			{Aharony}},\ }\href@noop {} {\emph {\bibinfo {title} {Introduction to
				percolation theory}}}\ (\bibinfo  {publisher} {CRC press},\ \bibinfo {year}
	{2018})\BibitemShut {NoStop}%
	\bibitem [{\citenamefont {Chayes}\ and\ \citenamefont
		{Chayes}(1987)}]{chayes1987upper}%
	\BibitemOpen
	\bibfield  {author} {\bibinfo {author} {\bibfnamefont {J.}~\bibnamefont
			{Chayes}}\ and\ \bibinfo {author} {\bibfnamefont {L.}~\bibnamefont
			{Chayes}},\ } {\bibfield  {journal}
		{\bibinfo  {journal} {Communications in Mathematical Physics}\ }\textbf
		{\bibinfo {volume} {113}},\ \bibinfo {pages} {27} (\bibinfo {year}
		{1987})}\BibitemShut {NoStop}%
	\bibitem [{\citenamefont {Aizenman}\ and\ \citenamefont
		{Newman}(1984)}]{AizenmanNewman1984}%
	\BibitemOpen
	\bibfield  {author} {\bibinfo {author} {\bibfnamefont {M.}~\bibnamefont
			{Aizenman}}\ and\ \bibinfo {author} {\bibfnamefont {C.~M.}\ \bibnamefont
			{Newman}},\ }{\bibfield  {journal}
		{\bibinfo  {journal} {Journal of Statistical Physics}\ }\textbf {\bibinfo
			{volume} {36}},\ \bibinfo {pages} {107} (\bibinfo {year} {1984})}\BibitemShut
	{NoStop}%
	\bibitem [{\citenamefont {Hara}\ and\ \citenamefont
		{Slade}(1990)}]{HaraSlade1990}%
	\BibitemOpen
	\bibfield  {author} {\bibinfo {author} {\bibfnamefont {T.}~\bibnamefont
			{Hara}}\ and\ \bibinfo {author} {\bibfnamefont {G.}~\bibnamefont {Slade}},\
	} {\bibfield  {journal} {\bibinfo
			{journal} {Communications in Mathematical Physics}\ }\textbf {\bibinfo
			{volume} {128}},\ \bibinfo {pages} {333–391} (\bibinfo {year}
		{1990})}\BibitemShut {NoStop}%
	\bibitem [{\citenamefont {Herega}(2015)}]{Herega2015}%
	\BibitemOpen
	\bibfield  {author} {\bibinfo {author} {\bibfnamefont {A.}~\bibnamefont
			{Herega}},\ } {\bibfield
		{journal} {\bibinfo  {journal} {Journal of Materials Science and Engineering
				A}\ }\textbf {\bibinfo {volume} {5}},\ \bibinfo {pages} {409} (\bibinfo
		{year} {2015})}\BibitemShut {NoStop}%
	\bibitem [{\citenamefont {Hopfield}(1982)}]{Hopfield1982}%
	\BibitemOpen
	\bibfield  {author} {\bibinfo {author} {\bibfnamefont {J.~J.}\ \bibnamefont
			{Hopfield}},\ } {\bibfield  {journal}
		{\bibinfo  {journal} {Proceedings of the National Academy of Sciences}\
		}\textbf {\bibinfo {volume} {79}},\ \bibinfo {pages} {2554} (\bibinfo {year}
		{1982})}\BibitemShut {NoStop}%
	\bibitem [{\citenamefont {Buldyrev}\ \emph {et~al.}(2010)\citenamefont
		{Buldyrev}, \citenamefont {Parshani}, \citenamefont {Paul}, \citenamefont
		{Eugene},\ and\ \citenamefont {Havlin}}]{SergeyStanley2010}%
	\BibitemOpen
	\bibfield  {author} {\bibinfo {author} {\bibfnamefont {S.~V.}\ \bibnamefont
			{Buldyrev}}, \bibinfo {author} {\bibfnamefont {R.}~\bibnamefont {Parshani}},
		\bibinfo {author} {\bibfnamefont {G.}~\bibnamefont {Paul}}, \bibinfo {author}
		{\bibfnamefont {S.~H.}\ \bibnamefont {Eugene}}, \ and\ \bibinfo {author}
		{\bibfnamefont {S.}~\bibnamefont {Havlin}},\ } {\bibfield  {journal} {\bibinfo  {journal} {Nature}\
		}\textbf {\bibinfo {volume} {464}},\ \bibinfo {pages} {1025–1028} (\bibinfo
		{year} {2010})}\BibitemShut {NoStop}%
	\bibitem [{\citenamefont {Mello}\ \emph {et~al.}(2021)\citenamefont {Mello},
		\citenamefont {Squillante}, \citenamefont {Gomes}, \citenamefont
		{Seridonio},\ and\ \citenamefont {de~Souza}}]{Mello2021}%
	\BibitemOpen
	\bibfield  {author} {\bibinfo {author} {\bibfnamefont {I.~F.}\ \bibnamefont
			{Mello}}, \bibinfo {author} {\bibfnamefont {L.}~\bibnamefont {Squillante}},
		\bibinfo {author} {\bibfnamefont {G.~O.}\ \bibnamefont {Gomes}}, \bibinfo
		{author} {\bibfnamefont {A.~C.}\ \bibnamefont {Seridonio}}, \ and\ \bibinfo
		{author} {\bibfnamefont {M.}~\bibnamefont {de~Souza}},\ } {\bibfield  {journal} {\bibinfo  {journal}
			{Physica A: Statistical Mechanics and its Applications}\ }\textbf {\bibinfo
			{volume} {573}},\ \bibinfo {pages} {125963} (\bibinfo {year}
		{2021})}\BibitemShut {NoStop}%
	\bibitem [{\citenamefont {Zeng}\ \emph {et~al.}(2020)\citenamefont {Zeng},
		\citenamefont {Gao}, \citenamefont {Shekhtman}, \citenamefont {Guo},
		\citenamefont {Lv}, \citenamefont {Wu}, \citenamefont {Liu}, \citenamefont
		{Levy}, \citenamefont {Li}, \citenamefont {Gao}, \citenamefont {Stanley},\
		and\ \citenamefont {Havlin}}]{ZengStanley2020}%
	\BibitemOpen
	\bibfield  {author} {\bibinfo {author} {\bibfnamefont {G.}~\bibnamefont
			{Zeng}}, \bibinfo {author} {\bibfnamefont {J.}~\bibnamefont {Gao}}, \bibinfo
		{author} {\bibfnamefont {L.}~\bibnamefont {Shekhtman}}, \bibinfo {author}
		{\bibfnamefont {S.}~\bibnamefont {Guo}}, \bibinfo {author} {\bibfnamefont
			{W.}~\bibnamefont {Lv}}, \bibinfo {author} {\bibfnamefont {J.}~\bibnamefont
			{Wu}}, \bibinfo {author} {\bibfnamefont {H.}~\bibnamefont {Liu}}, \bibinfo
		{author} {\bibfnamefont {O.}~\bibnamefont {Levy}}, \bibinfo {author}
		{\bibfnamefont {D.}~\bibnamefont {Li}}, \bibinfo {author} {\bibfnamefont
			{Z.}~\bibnamefont {Gao}}, \bibinfo {author} {\bibfnamefont {H.~E.}\
			\bibnamefont {Stanley}}, \ and\ \bibinfo {author} {\bibfnamefont
			{S.}~\bibnamefont {Havlin}},\ }
	{\bibfield  {journal} {\bibinfo  {journal} {Proceedings of the National
				Academy of Sciences}\ }\textbf {\bibinfo {volume} {117}},\ \bibinfo {pages}
		{17528} (\bibinfo {year} {2020})}\BibitemShut {NoStop}%
	\bibitem [{\citenamefont {Ma}\ \emph {et~al.}(2019)\citenamefont {Ma},
		\citenamefont {Sudakov}, \citenamefont {Strong},\ and\ \citenamefont
		{Golden}}]{Ma2019}%
	\BibitemOpen
	\bibfield  {author} {\bibinfo {author} {\bibfnamefont {Y.-P.}\ \bibnamefont
			{Ma}}, \bibinfo {author} {\bibfnamefont {I.}~\bibnamefont {Sudakov}},
		\bibinfo {author} {\bibfnamefont {C.}~\bibnamefont {Strong}}, \ and\ \bibinfo
		{author} {\bibfnamefont {K.~M.}\ \bibnamefont {Golden}},\ } {\bibfield  {journal} {\bibinfo  {journal} {New
				Journal of Physics}\ }\textbf {\bibinfo {volume} {21}},\ \bibinfo {pages}
		{063029} (\bibinfo {year} {2019})}\BibitemShut {NoStop}%
	\bibitem [{\citenamefont {Brunk}\ and\ \citenamefont
		{Twarock}(2021)}]{BrunkTwarock2021}%
	\BibitemOpen
	\bibfield  {author} {\bibinfo {author} {\bibfnamefont {N.~E.}\ \bibnamefont
			{Brunk}}\ and\ \bibinfo {author} {\bibfnamefont {R.}~\bibnamefont
			{Twarock}},\ } {\bibfield  {journal}
		{\bibinfo  {journal} {ACS Nano}\ }\textbf {\bibinfo {volume} {15}},\ \bibinfo
		{pages} {12988–12995} (\bibinfo {year} {2021})}\BibitemShut {NoStop}%
	\bibitem [{\citenamefont {Zhang}\ \emph {et~al.}(2019)\citenamefont {Zhang},
		\citenamefont {Zeng}, \citenamefont {Li}, \citenamefont {Huang},
		\citenamefont {Stanley},\ and\ \citenamefont {Havlin}}]{ZhangStanley2019}%
	\BibitemOpen
	\bibfield  {author} {\bibinfo {author} {\bibfnamefont {L.}~\bibnamefont
			{Zhang}}, \bibinfo {author} {\bibfnamefont {G.}~\bibnamefont {Zeng}},
		\bibinfo {author} {\bibfnamefont {D.}~\bibnamefont {Li}}, \bibinfo {author}
		{\bibfnamefont {H.-J.}\ \bibnamefont {Huang}}, \bibinfo {author}
		{\bibfnamefont {H.~E.}\ \bibnamefont {Stanley}}, \ and\ \bibinfo {author}
		{\bibfnamefont {S.}~\bibnamefont {Havlin}},\ } {\bibfield  {journal} {\bibinfo  {journal}
			{Proceedings of the National Academy of Sciences}\ }\textbf {\bibinfo
			{volume} {116}},\ \bibinfo {pages} {8673} (\bibinfo {year}
		{2019})}\BibitemShut {NoStop}%
	\bibitem [{\citenamefont {Grimmett}(2006)}]{Grimmett2006}%
	\BibitemOpen
	\bibfield  {author} {\bibinfo {author} {\bibfnamefont {G.}~\bibnamefont
			{Grimmett}},\ }
	{\emph {\bibinfo {title} {The Random-Cluster Model}}},\ Grundlehren der
	mathematischen Wissenschaften\ (\bibinfo  {publisher} {Springer Berlin
		Heidelberg},\ \bibinfo {year} {2006})\BibitemShut {NoStop}%
	\bibitem [{\citenamefont {Wu}(1982)}]{Wu1982}%
	\BibitemOpen
	\bibfield  {author} {\bibinfo {author} {\bibfnamefont {F.~Y.}\ \bibnamefont
			{Wu}},\ } {\bibfield  {journal}
		{\bibinfo  {journal} {Reviews of Modern Physics}\ }\textbf {\bibinfo {volume}
			{54}},\ \bibinfo {pages} {235} (\bibinfo {year} {1982})}\BibitemShut
	{NoStop}%
	\bibitem [{\citenamefont {Swendsen}\ and\ \citenamefont
		{Wang}(1987)}]{swendsen1987nonuniversal}%
	\BibitemOpen
	\bibfield  {author} {\bibinfo {author} {\bibfnamefont {R.~H.}\ \bibnamefont
			{Swendsen}}\ and\ \bibinfo {author} {\bibfnamefont {J.-S.}\ \bibnamefont
			{Wang}},\ } {\bibfield  {journal}
		{\bibinfo  {journal} {Physical Review Letters}\ }\textbf {\bibinfo {volume}
			{58}},\ \bibinfo {pages} {86} (\bibinfo {year} {1987})}\BibitemShut {NoStop}%
	\bibitem [{\citenamefont {Wolff}(1989)}]{wolff1989collective}%
	\BibitemOpen
	\bibfield  {author} {\bibinfo {author} {\bibfnamefont {U.}~\bibnamefont
			{Wolff}},\ } {\bibfield  {journal}
		{\bibinfo  {journal} {Physical Review Letters}\ }\textbf {\bibinfo {volume}
			{62}},\ \bibinfo {pages} {361} (\bibinfo {year} {1989})}\BibitemShut
	{NoStop}%
	\bibitem [{\citenamefont {Chayes}\ and\ \citenamefont
		{Machta}(1998)}]{chayes1998graphicala}%
	\BibitemOpen
	\bibfield  {author} {\bibinfo {author} {\bibfnamefont {L.}~\bibnamefont
			{Chayes}}\ and\ \bibinfo {author} {\bibfnamefont {J.}~\bibnamefont
			{Machta}},\ } {\bibfield
		{journal} {\bibinfo  {journal} {Physica A: Statistical Mechanics and its
				Applications}\ }\textbf {\bibinfo {volume} {254}},\ \bibinfo {pages} {477}
		(\bibinfo {year} {1998})}\BibitemShut {NoStop}%
	\bibitem [{\citenamefont {Zhang}\ \emph {et~al.}(2020)\citenamefont {Zhang},
		\citenamefont {Michel}, \citenamefont {El{\c c}i},\ and\ \citenamefont
		{Deng}}]{zhang2020loop}%
	\BibitemOpen
	\bibfield  {author} {\bibinfo {author} {\bibfnamefont {L.}~\bibnamefont
			{Zhang}}, \bibinfo {author} {\bibfnamefont {M.}~\bibnamefont {Michel}},
		\bibinfo {author} {\bibfnamefont {E.~M.}\ \bibnamefont {El{\c c}i}}, \ and\
		\bibinfo {author} {\bibfnamefont {Y.}~\bibnamefont {Deng}},\ } {\bibfield  {journal} {\bibinfo  {journal}
			{Physical Review Letters}\ }\textbf {\bibinfo {volume} {125}},\ \bibinfo
		{pages} {200603} (\bibinfo {year} {2020})}\BibitemShut {NoStop}%
	\bibitem [{Note2()}]{Note2}%
	\BibitemOpen
	\bibinfo {note} {A complete graph with $V$ vertices is a graph in which each
		vertex is connected to all others.}\BibitemShut {Stop}%
	\bibitem [{\citenamefont {Chayes}\ \emph {et~al.}(1999)\citenamefont {Chayes},
		\citenamefont {Coniglio}, \citenamefont {Machta},\ and\ \citenamefont
		{Shtengel}}]{chayes1999meanfield}%
	\BibitemOpen
	\bibfield  {author} {\bibinfo {author} {\bibfnamefont {L.}~\bibnamefont
			{Chayes}}, \bibinfo {author} {\bibfnamefont {A.}~\bibnamefont {Coniglio}},
		\bibinfo {author} {\bibfnamefont {J.}~\bibnamefont {Machta}}, \ and\ \bibinfo
		{author} {\bibfnamefont {K.}~\bibnamefont {Shtengel}},\ } {\bibfield
		{journal} {\bibinfo  {journal} {Journal of Statistical Physics}\ }\textbf
		{\bibinfo {volume} {94}},\ \bibinfo {pages} {53} (\bibinfo {year}
		{1999})}\BibitemShut {NoStop}%
	\bibitem [{\citenamefont {Luczak}\ and\ \citenamefont
		{\L~uczak}(2006)}]{LuczakLuczak2008}%
	\BibitemOpen
	\bibfield  {author} {\bibinfo {author} {\bibfnamefont {M.}~\bibnamefont
			{Luczak}}\ and\ \bibinfo {author} {\bibfnamefont {T.}~\bibnamefont
			{\L~uczak}},\ } {\bibfield  {journal}
		{\bibinfo  {journal} {Random Structures \& Algorithms}\ }\textbf {\bibinfo
			{volume} {28}},\ \bibinfo {pages} {215} (\bibinfo {year} {2006})}\BibitemShut
	{NoStop}%
	\bibitem [{\citenamefont {Fang}\ \emph
		{et~al.}(2021{\natexlab{a}})\citenamefont {Fang}, \citenamefont {Zhou},\ and\
		\citenamefont {Deng}}]{FangZhouDeng2021}%
	\BibitemOpen
	\bibfield  {author} {\bibinfo {author} {\bibfnamefont {S.}~\bibnamefont
			{Fang}}, \bibinfo {author} {\bibfnamefont {Z.}~\bibnamefont {Zhou}}, \ and\
		\bibinfo {author} {\bibfnamefont {Y.}~\bibnamefont {Deng}},\ } {\bibfield  {journal} {\bibinfo  {journal}
			{Physical Review E}\ }\textbf {\bibinfo {volume} {103}},\ \bibinfo {pages}
		{012102} (\bibinfo {year} {2021}{\natexlab{a}})}\BibitemShut {NoStop}%
	\bibitem [{\citenamefont {Fang}\ \emph {et~al.}(2020)\citenamefont {Fang},
		\citenamefont {Grimm}, \citenamefont {Zhou},\ and\ \citenamefont
		{Deng}}]{FangGrimmZhouDeng2020}%
	\BibitemOpen
	\bibfield  {author} {\bibinfo {author} {\bibfnamefont {S.}~\bibnamefont
			{Fang}}, \bibinfo {author} {\bibfnamefont {J.}~\bibnamefont {Grimm}},
		\bibinfo {author} {\bibfnamefont {Z.}~\bibnamefont {Zhou}}, \ and\ \bibinfo
		{author} {\bibfnamefont {Y.}~\bibnamefont {Deng}},\ } {\bibfield  {journal} {\bibinfo  {journal}
			{Physical Review E}\ }\textbf {\bibinfo {volume} {102}},\ \bibinfo {pages}
		{022125} (\bibinfo {year} {2020})}\BibitemShut {NoStop}%
	\bibitem [{\citenamefont {Lv}\ \emph {et~al.}(2020)\citenamefont {Lv},
		\citenamefont {Xu}, \citenamefont {Sun}, \citenamefont {Chen},\ and\
		\citenamefont {Deng}}]{Lv2019Two}%
	\BibitemOpen
	\bibfield  {author} {\bibinfo {author} {\bibfnamefont {J.-P.}\ \bibnamefont
			{Lv}}, \bibinfo {author} {\bibfnamefont {W.}~\bibnamefont {Xu}}, \bibinfo
		{author} {\bibfnamefont {Y.}~\bibnamefont {Sun}}, \bibinfo {author}
		{\bibfnamefont {K.}~\bibnamefont {Chen}}, \ and\ \bibinfo {author}
		{\bibfnamefont {Y.}~\bibnamefont {Deng}},\ } {\bibfield  {journal} {\bibinfo  {journal} {National
				Science Review}\ } (\bibinfo {year} {2020}),\
		10.1093/nsr/nwaa212}\BibitemShut {NoStop}%
	\bibitem [{\citenamefont {Lundow}\ and\ \citenamefont
		{Markstr{\"o}m}(2015)}]{lundow2015discontinuity}%
	\BibitemOpen
	\bibfield  {author} {\bibinfo {author} {\bibfnamefont {P.}~\bibnamefont
			{Lundow}}\ and\ \bibinfo {author} {\bibfnamefont {K.}~\bibnamefont
			{Markstr{\"o}m}},\ }
	{\bibfield  {journal} {\bibinfo  {journal} {Nuclear Physics B}\ }\textbf
		{\bibinfo {volume} {895}},\ \bibinfo {pages} {305} (\bibinfo {year}
		{2015})}\BibitemShut {NoStop}%
	\bibitem [{\citenamefont {Mertens}\ and\ \citenamefont
		{Moore}(2018)}]{MertensMoore2018}%
	\BibitemOpen
	\bibfield  {author} {\bibinfo {author} {\bibfnamefont {S.}~\bibnamefont
			{Mertens}}\ and\ \bibinfo {author} {\bibfnamefont {C.}~\bibnamefont
			{Moore}},\ } {\bibfield  {journal}
		{\bibinfo  {journal} {Physical Review E}\ }\textbf {\bibinfo {volume} {98}},\
		\bibinfo {pages} {022120} (\bibinfo {year} {2018})}\BibitemShut {NoStop}%
	\bibitem [{\citenamefont {Wittmann}\ and\ \citenamefont
		{Young}(2014)}]{wittmann2014finitesize}%
	\BibitemOpen
	\bibfield  {author} {\bibinfo {author} {\bibfnamefont {M.}~\bibnamefont
			{Wittmann}}\ and\ \bibinfo {author} {\bibfnamefont {A.~P.}\ \bibnamefont
			{Young}},\ } {\bibfield  {journal}
		{\bibinfo  {journal} {Physical Review E}\ }\textbf {\bibinfo {volume} {90}},\
		\bibinfo {pages} {062137} (\bibinfo {year} {2014})}\BibitemShut {NoStop}%
	\bibitem [{\citenamefont {{Flores-Sola}}\ \emph {et~al.}(2016)\citenamefont
		{{Flores-Sola}}, \citenamefont {Berche}, \citenamefont {Kenna},\ and\
		\citenamefont {Weigel}}]{flores-sola2016role}%
	\BibitemOpen
	\bibfield  {author} {\bibinfo {author} {\bibfnamefont {E.}~\bibnamefont
			{{Flores-Sola}}}, \bibinfo {author} {\bibfnamefont {B.}~\bibnamefont
			{Berche}}, \bibinfo {author} {\bibfnamefont {R.}~\bibnamefont {Kenna}}, \
		and\ \bibinfo {author} {\bibfnamefont {M.}~\bibnamefont {Weigel}},\ } {\bibfield  {journal} {\bibinfo
			{journal} {Physical Review Letters}\ }\textbf {\bibinfo {volume} {116}},\
		\bibinfo {pages} {115701} (\bibinfo {year} {2016})}\BibitemShut {NoStop}%
	\bibitem [{\citenamefont {Grimm}\ \emph {et~al.}(2017)\citenamefont {Grimm},
		\citenamefont {El{\c c}i}, \citenamefont {Zhou}, \citenamefont {Garoni},\
		and\ \citenamefont {Deng}}]{grimm2017geometric}%
	\BibitemOpen
	\bibfield  {author} {\bibinfo {author} {\bibfnamefont {J.}~\bibnamefont
			{Grimm}}, \bibinfo {author} {\bibfnamefont {E.~M.}\ \bibnamefont {El{\c
					c}i}}, \bibinfo {author} {\bibfnamefont {Z.}~\bibnamefont {Zhou}}, \bibinfo
		{author} {\bibfnamefont {T.~M.}\ \bibnamefont {Garoni}}, \ and\ \bibinfo
		{author} {\bibfnamefont {Y.}~\bibnamefont {Deng}},\ } {\bibfield  {journal} {\bibinfo  {journal}
			{Physical Review Letters}\ }\textbf {\bibinfo {volume} {118}},\ \bibinfo
		{pages} {115701} (\bibinfo {year} {2017})}\BibitemShut {NoStop}%
	\bibitem [{\citenamefont {Zhou}\ \emph {et~al.}(2018)\citenamefont {Zhou},
		\citenamefont {Grimm}, \citenamefont {Fang}, \citenamefont {Deng},\ and\
		\citenamefont {Garoni}}]{zhou2018randomlength}%
	\BibitemOpen
	\bibfield  {author} {\bibinfo {author} {\bibfnamefont {Z.}~\bibnamefont
			{Zhou}}, \bibinfo {author} {\bibfnamefont {J.}~\bibnamefont {Grimm}},
		\bibinfo {author} {\bibfnamefont {S.}~\bibnamefont {Fang}}, \bibinfo {author}
		{\bibfnamefont {Y.}~\bibnamefont {Deng}}, \ and\ \bibinfo {author}
		{\bibfnamefont {T.~M.}\ \bibnamefont {Garoni}},\ } {\bibfield  {journal} {\bibinfo  {journal}
			{Physical Review Letters}\ }\textbf {\bibinfo {volume} {121}},\ \bibinfo
		{pages} {185701} (\bibinfo {year} {2018})}\BibitemShut {NoStop}%
	\bibitem [{\citenamefont {Fang}\ \emph
		{et~al.}(2021{\natexlab{b}})\citenamefont {Fang}, \citenamefont {Deng},\ and\
		\citenamefont {Zhou}}]{FangDengZhou2021}%
	\BibitemOpen
	\bibfield  {author} {\bibinfo {author} {\bibfnamefont {S.}~\bibnamefont
			{Fang}}, \bibinfo {author} {\bibfnamefont {Y.}~\bibnamefont {Deng}}, \ and\
		\bibinfo {author} {\bibfnamefont {Z.}~\bibnamefont {Zhou}},\ } {\bibfield  {journal} {\bibinfo  {journal}
			{Physical Review E}\ }\textbf {\bibinfo {volume} {104}},\ \bibinfo {pages}
		{064108} (\bibinfo {year} {2021}{\natexlab{b}})}\BibitemShut {NoStop}%
	\bibitem [{\citenamefont {{V. Papathanakos}}(2006)}]{Papathanakos2006}%
	\BibitemOpen
	\bibfield  {author} {\bibinfo {author} {\bibnamefont {{V. Papathanakos}}},\
	}\emph {\bibinfo {title} {{Finite-Size Effects in High-Dimensional
				Statistical Mechanical Systems: The Ising Model With Periodic Boundary
				Conditions}}},\ \href@noop {} {Ph.D. thesis},\ \bibinfo  {school} {{Princeton
			University}}, \bibinfo {address} {{Princeton, New Jersey}} (\bibinfo {year}
	{2006})\BibitemShut {NoStop}%
	\bibitem [{\citenamefont {{J. Grimm, E. M. El\c{c}i, Z. Zhou, T. M. Garoni and
				Y. Deng}}(2017)}]{GrimmElciZhouGaroniDeng2017}%
	\BibitemOpen
	\bibfield  {author} {\bibinfo {author} {\bibnamefont {{J. Grimm, E. M.
					El\c{c}i, Z. Zhou, T. M. Garoni and Y. Deng}}},\ } {\bibfield  {journal} {\bibinfo  {journal}
			{{Physical Review Letters}}\ }\textbf {\bibinfo {volume} {118}},\ \bibinfo
		{pages} {115701} (\bibinfo {year} {2017})}\BibitemShut {NoStop}%
	\bibitem [{\citenamefont {{Z. Zhou, J. Grimm, S. Fang, Y. Deng, and T. M.
				Garoni}}(2018)}]{ZhouGrimmFangDengGaroni2018}%
	\BibitemOpen
	\bibfield  {author} {\bibinfo {author} {\bibnamefont {{Z. Zhou, J. Grimm, S.
					Fang, Y. Deng, and T. M. Garoni}}},\ } {\bibfield  {journal} {\bibinfo  {journal}
			{{Physical Review Letters}}\ }\textbf {\bibinfo {volume} {121}},\ \bibinfo
		{pages} {185701} (\bibinfo {year} {2018})}\BibitemShut {NoStop}%
	\bibitem [{\citenamefont {Bollob{\'a}s}\ \emph {et~al.}(1996)\citenamefont
		{Bollob{\'a}s}, \citenamefont {Grimmett},\ and\ \citenamefont
		{Janson}}]{BollobasGrimmettJanson1996}%
	\BibitemOpen
	\bibfield  {author} {\bibinfo {author} {\bibfnamefont {B.}~\bibnamefont
			{Bollob{\'a}s}}, \bibinfo {author} {\bibfnamefont {G.}~\bibnamefont
			{Grimmett}}, \ and\ \bibinfo {author} {\bibfnamefont {S.}~\bibnamefont
			{Janson}},\ } {\bibfield  {journal}
		{\bibinfo  {journal} {Probability Theory and Related Fields}\ }\textbf
		{\bibinfo {volume} {104}},\ \bibinfo {pages} {283} (\bibinfo {year}
		{1996})}\BibitemShut {NoStop}%
	\bibitem [{\citenamefont {Kenna}(2013)}]{kenna2013universal}%
	\BibitemOpen
	\bibfield  {author} {\bibinfo {author} {\bibfnamefont {R.}~\bibnamefont
			{Kenna}},\ }in\ \href@noop {} {\emph {\bibinfo {booktitle} {Order, Disorder
				and Criticality: Advanced Problems of Phase Transition Theory Volume 3}}}\
	(\bibinfo  {publisher} {World Scientific},\ \bibinfo {year} {2013})\ pp.\
	\bibinfo {pages} {1--46}\BibitemShut {NoStop}%
	\bibitem [{\citenamefont {Kenna}(2004)}]{kenna2004finite}%
	\BibitemOpen
	\bibfield  {author} {\bibinfo {author} {\bibfnamefont {R.}~\bibnamefont
			{Kenna}},\ }\href@noop {} {\bibfield  {journal} {\bibinfo  {journal} {Nuclear
				Physics B}\ }\textbf {\bibinfo {volume} {691}},\ \bibinfo {pages} {292 }
		(\bibinfo {year} {2004})}\BibitemShut {NoStop}%
	\bibitem [{\citenamefont {Aharony}\ \emph {et~al.}(1984)\citenamefont
		{Aharony}, \citenamefont {Gefen},\ and\ \citenamefont
		{Kapitulnik}}]{aharony1984scaling}%
	\BibitemOpen
	\bibfield  {author} {\bibinfo {author} {\bibfnamefont {A.}~\bibnamefont
			{Aharony}}, \bibinfo {author} {\bibfnamefont {Y.}~\bibnamefont {Gefen}}, \
		and\ \bibinfo {author} {\bibfnamefont {A.}~\bibnamefont {Kapitulnik}},\
	} {\bibfield  {journal} {\bibinfo
			{journal} {Journal of Physics A: Mathematical and General}\ }\textbf
		{\bibinfo {volume} {17}},\ \bibinfo {pages} {L197} (\bibinfo {year}
		{1984})}\BibitemShut {NoStop}%
	\bibitem [{\citenamefont {Nahum}\ \emph {et~al.}(2013)\citenamefont {Nahum},
		\citenamefont {Chalker}, \citenamefont {Serna}, \citenamefont {Ortuno},\ and\
		\citenamefont {Somoza}}]{nahum2013phase}%
	\BibitemOpen
	\bibfield  {author} {\bibinfo {author} {\bibfnamefont {A.}~\bibnamefont
			{Nahum}}, \bibinfo {author} {\bibfnamefont {J.}~\bibnamefont {Chalker}},
		\bibinfo {author} {\bibfnamefont {P.}~\bibnamefont {Serna}}, \bibinfo
		{author} {\bibfnamefont {M.}~\bibnamefont {Ortuno}}, \ and\ \bibinfo {author}
		{\bibfnamefont {A.}~\bibnamefont {Somoza}},\ } {\bibfield  {journal} {\bibinfo  {journal}
			{Physical Review B}\ }\textbf {\bibinfo {volume} {88}},\ \bibinfo {pages}
		{134411} (\bibinfo {year} {2013})}\BibitemShut {NoStop}%
\end{thebibliography}
%

\end{document}